\documentstyle[psfig]{l-aa}

\begin{document}
 \title{Two-photon annihilation in the pair formation cascades in pulsar 
	polar caps}

 \author{Bing Zhang\inst{1,2}, G. J. Qiao\inst{3,1,2}}
%%%% \offprints{B. Zhang.}
 \offprints{B. Zhang.\\ $^\dag$ BAC is jointly sponsored by the
 	Chinese Academy of Sciences and Peking University}
 \institute{Department of Geophysics, Peking University, Beijing, 100871,
              P. R. China, 
	\and  Beijing Astrophysics Center (BAC)$^\dag$, 
              Beijing, 100871, P. R. China,
	\and  CCAST, P.O. Box 8730, Beijing 100080, P. R. China}
 \thesaurus{00.00.0, 00.00.0}
 \date{Received date ; accepted date}
 \maketitle
\markboth{B.Zhang, \& G.J.Qiao: Two-photon annihilation in pulsar polar caps}{}

%%\footnotetext{BAC is jointly sponsored by the Chinese Academy of Sciences and
%%              Peking University}

%-----------------------------------------------------------------------

\begin{abstract}  

The importance of the photon-photon pair production process ($\gamma+
\gamma^{\prime}\rightarrow e^{+}+e^{-}$) to form pair production cascades
 in pulsar polar caps is investigated within the framework of the
 Ruderman-Sutherland vacuum gap model. It is found that this process is
 unimportant if the polar caps are not hot enough, but will play a
 non-negligible role in the pair formation cascades when the polar cap
 temperatures are in excess of the critical temperatures, $T_{cri}$,
 which are around $4\times 10^6K$ when $P=0.1$s and will slowly
 increase with increasing periods. Compared with the $\gamma-B$
 process, it is found that the two-photon annihilation process may ignite
 a central spark near the magnetic pole, where $\gamma-B$ sparks can
 not be formed due to the local weak curvatures. This central spark is
 large if the gap is dominated by the ``resonant ICS mode''. The
 possible connection of these central sparks with the observed pulsar
 ``core'' emission components is discussed.

\end{abstract}


\section{Introduction}

 Recent X-ray observations and data analyses show strong evidence of thermal
 emission components from the whole surfaces as well as the hot polar caps of
 some pulsars (e.g. \"Ogelman 1995; Greiveldinger et al. 1996; Wang \& Halpern
 1997; Becker \& Tr\"umper 1997). The presence of these thermal photons near
 the neutron star (especially near the polar caps) may bring various
 new physical effects which have not been fully investigated before. One
 important process which has stimulated interest is the inverse
 Compton scattering (ICS) processing of these photons by the high energy
 particles. Recent studies show that the ICS process is not only a significant
 energy loss (Xia et al. 1985; Daugherty \& Harding 1989; Dermer 1990; Chang
 1995; Sturner 1995) and radiation (Sturner \& Dermer 1994) mechanism, but is
 also the dominant mechanism to ignite pair production cascades near the pulsar
 polar caps and hence, to limit the parameters of the pulsar inner accelerators
 (Zhang \& Qiao 1996, hereafter ZQ96; Qiao \& Zhang 1996; Luo 1996; Zhang et
 al. 1997a, hereafter ZQLH97a; Zhang et al. 1997b, hereafter ZQH97b; Harding
 \& Muslimov 1998, hereafter HM98). Within the framework of the
 Ruderman-Sutherland (1975, hereafter RS75) vacuum gap model, ZQ96 found that
 the ICS-induced $\gamma-B$ process is the most important mechanism to cause
 an inner gap breakdown. The two characteristic frequencies in the
 up-scattered ICS spectra lead to two different modes of the inner gaps,
 which are dominant within different temperature regimes (ZQLH97a, ZQH97b).

 The mean free path of pair production is the key parameter to study pulsar
 polar cap physics, since it is the relevant quantity
 to determine the length of
 the inner accelerator. It is necessary to compare the relative importance of
 the various pair formation processes in the pulsar polar cap region. {\em A
 process with shorter pair formation mean free path will be more likely to
 dominate pair formation cascades in the inner gaps.} For example, for the
 $\gamma-B$ process, the ICS-induced pair formation mean free path is shorter
 than the curvature radiation (CR)-induced one, so the CR mode is usually
 suppressed. For the ICS process, when the polar cap temperature is high
 enough so that the scatterings off the soft photons near the peak
 of the Planck spectrum
 become important, this so-called ``thermal-peak ICS mode'' will suppress the
 otherwise dominant ``resonant ICS mode'', which is mainly the contribution of
 the resonant ICS process. This is also because the mean free path of the
 ``thermal'' ICS mode is shorter than that of the ``resonant'' one (ZQLH97a,
 ZQH97b).

 Another process which might also be important with the presence of the hot
 thermal photon fields in the neutron star vicinity is the photon-photon pair
 production or the two-photon annihilation. The two ingredients which take part
 in the process are: the soft thermal X-rays and their inverse Compton
 $\gamma$-rays. In this paper, we will explicitly study the possible
 importance of the $\gamma-\gamma$ process and compare its mean free
 path with that of the $\gamma-B$
 process. The method to calculate the mean free path of the two-photon
 annihilation in the polar caps is described in
 Sect.2. A comparison between the relative importance of the $\gamma-\gamma$
 and the $\gamma-B$ processes in a dipolar magnetic field configuration is
 presented in Sect.3. The possible connection of the $\gamma- \gamma$-induced
 central sparks with the pulsar ``core'' emission components is discussed in
 Sect.4.

\section{Two-photon annihilation and polar gap sparking}

 The threshold condition of two-photon annihilation for a high-energy photon
 (energy $E$) and a low-energy photon (energy $\epsilon$) colliding with an
 incident angle of $\cos^{-1} \mu_c$ is
 $$\epsilon E (1-\mu_c)\geq 2(m_0c^2)^2, \eqno(1)$$
 where $m_0c^2$ is the static energy of the electron/positron. The high energy
 $\gamma$ photons produced by the ICS process by the quasi-monoenergetic
 electrons/positrons over the semi-isotropic thermal photons are found to
 have two characteristic energies, the relative importance of which is
 determined by the polar cap temperature (ZQLH97a). The first characteristic
 energy is the contribution of the ``resonant'' scatterings, in which the
 scattering cross sections are tremendously enhanced due to the rapid
 transitions between the ground and the first Landau levels of the electrons
 in strong magnetic fields. Various resonant scatterings over photons with
 different energies coming from different angles all contribute to this sharp
 characteristic energy 
 $$E_{res}=2\gamma \hbar \omega_B=2\gamma\hbar{eB\over m_0c}\simeq 2.3\times
 10^3 \gamma_5 B_{12} {\rm MeV}, \eqno(2)$$
 where $\gamma=10^5\gamma_5$ is the Lorentz factor of the electrons/positrons,
 $\omega_B=eB/m_0c$ is the cyclotron frequency in a magnetic field $B=10^{12}
 B_{12}{\rm G}$. The second broad peak in the ICS spectrum is the contribution
 of the scatterings over the soft thermal X-rays near the Planck peak, with
 the characteristic energy
 $$E_{th}\sim \gamma^2 \cdot 2.82kT\simeq 2.4\times 10^4 \gamma_4^2 T_6
 {\rm MeV}, \eqno(3)$$
 where $T=10^6T_6{\rm K}$ is the polar cap temperature. This second ``thermal
 mode'' is only important when the polar cap temperature is high enough 
 (beyond the ``critical temperature'', see details in ZQLH97a). For a
 blackbody-like low-frequency photon field, we adopt
 $$\epsilon\sim 2.82kT\simeq 240T_6{\rm eV}. \eqno(4)$$
 From Eqs.[2-4] we find that the threshold condition (1) is easily 
 satisfied, since the scatterings have a wide range distribution of $\mu_c$.

 To calculate the mean free path of two-photon annihilation, we follow
 the approach of Gould \& Schr\'{e}der (1967, hereafter GS67), but generalize
 it to be able to treat the case of anisotropic collisions. 
 Assuming that the soft
 photon gas near the neutron star surface is blackbody-like (a discussion
 concerning this assumption is presented in the last section), then the
 absorption probability per unit path length is
 $${d\tau \over dx}(x)={\alpha^2 \over \pi\Lambda}({kT\over m_0c^2})^3
 f(\nu,x)\simeq 2.1\times 10^{-6}T_6^3 f(\nu,x){\rm cm^{-1}}, \eqno(5)$$
 with $\alpha=e^2/\hbar c\simeq 1/137$, $\Lambda=\hbar/m_0 c\simeq3.86\times
 10^{-11}{\rm cm}$, and $\nu=(m_0c^2)^2/(EkT)$. The function $f(\nu,x)$ reads
 $$f(\nu,x)=\nu^2\int_{{2\over 1-\mu_c(x)}\nu}^\infty (e^\epsilon -1)^{-1}
 \bar{\varphi}(\epsilon /\nu,x) d\epsilon \eqno(6)$$
 (GS67, their Eqs.[17,18]). Our generalization of the GS67 formalism is to
 multiply the lower-limit of integration of GS67 to a factor of $2/(1-\mu_c)$
 so as to be able to deal with the case of anisotropic collisions. Also the
 function $\bar{\varphi}(s_0,x)$ is re-defined as
 $$\bar{\varphi}(s_0,x)=\int_1^{{1-\mu_c(x) \over 2}s_0(\epsilon)}s \bar
 {\sigma}(s)ds, \eqno(7)$$
 with the upper limit of integration multiplied by a factor of $(1-\mu_c)/2$
 with respect to the GS67's result (their eq.[9]). Here $s_0=\epsilon E/m_0^2
 c^4$, and the integrand of eq.[7] can be deduced as
 $$s \bar{\sigma}(s)=(2+{2\over s}-{1\over s^2})\ln{1+(1-{1\over s})^{1/2}
 \over 1-(1-{1\over s})^{1/2}}-2(1-{1\over s})^{1/2}(1+{1\over s}) \eqno(8)$$
 according to the GS67 formalism (see their Eqs.[1,3,4,9]).

 It is of interests to calculate the shortest mean free path, so it is
 necessary to get the maximum value of the function $f(\nu)$. For an isotropic
 soft photon field, the minimum incident angle cosine is $\mu_c=-1$, which
 means that head-on collisions are important. In this case, Eqs.(7,8) are
 reduced to the GS67 results (their Eqs.[18,9]), and Max$[f(\nu)]\simeq 1$.
 For the case of pulsar polar cap region, however, the emission is directed
 radially outward, so that the photon gas is only semi-isotropic ($\mu_c =0$).
 Moreover, at higher altitudes, $\mu_c$ is even larger due to the finite size
 of the polar cap. In these cases, Max$[f(\nu)]$ cannot be that large. With
 variable $\mu_c$, it is not easy to get an analytic expression and
 asymptotic forms of eq.[6] similar to what GS67 did. We thus performed a 
 numerical
 calculation to get $\mu_c-{\rm Max}[f(\nu)]$ relation. The results are shown
 in Fig.1. A fairly good polynomial fit to the data gives
 $${\rm Max}[f(\nu)]=0.270-0.507\mu_c+0.237\mu_c^2, \eqno(9)$$
 which we will use directly in the further calculations. We see that 
 Max$[f(\nu)]
 \simeq1$ when $\mu_c=-1$ according to Eq.[9], which is in accordance to
 the GS67 result.

\begin{figure}
% \picplace{6.0 cm}
 \centerline{\psfig{figure=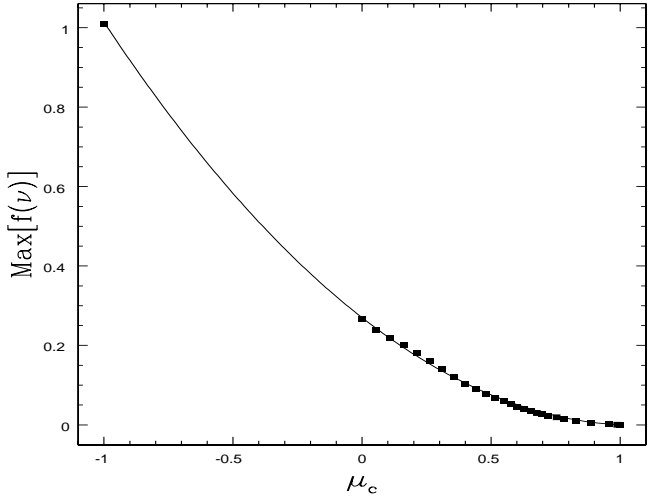,angle=0,height=7.0cm,width=9.0cm}}
 \caption[]{The numerical results and a polynomial fit (Eq.[9]) of the
 $\mu_c-{\rm Max}[f(\nu)]$ graph.}
\end{figure} 

 With the above mentioned preparation done, we can calculate
 the shortest mean free path of the two-photon annihilation process. This
 minimum mean free path (denoted by $l_{\gamma-\gamma,min}$) can be obtained
 through the relation
 $$\int_0^{l_{\gamma-\gamma,min}}({d\tau\over dx})_{max}dx$$
 $$=\int_0^{l_{\gamma-\gamma,min}}2.1\times 10^{-6}T_6^3 {\rm Max}[f(\nu,x)]
 dx=\tau=1, \eqno(10)$$
 where Max$f(\nu,x)$ is a function of $\mu_c$, and hence, of the height $x$
 from the surface.

 The last relationship which is necessary to calculate $l_{\gamma-\gamma, min}$
 is the $x$ dependence of $\mu_c$. First, it is worth noticing that there
 exists a thermal ``corona'' (Chang 1995) or ``atmosphere'' (e.g. Pavlov et al.
 1995) near the neutron star surface, the scale height of which is determined
 by $m_0gh\sim kT$, where $g\sim 10^{14}-10^{15}$cm s$^{-2}$ is the typical
 gravitational acceleration of a neutron star. Hence, this scale height is 
 $$h=x_0\simeq 820T_6 {\rm cm}. \eqno(11)$$
 Within this hot corona, the soft photon gas can be regarded as nearly 
 isotropic, since thermal equilibrium might be set up there. Beyond this hot
 corona, the photon gas begins to show strong anisotropy, so that $\mu_c$
 increases with $x$, while Max$[f(\nu)]$ decreases rapidly with $x$.
 For the configuration of a hot spot with the dimension of the polar cap,
 the $\mu_c-x$ relation is
 $$\mu_c=\left\{ \begin{array}{ll}
 -1, & x<x_0 \\
 (1-({R+x_0 \over R+x})^2)^{1/2}, & x_0\le x< x_{cri} \\
 {x-x_0 \over \sqrt{r_p^2+(x-x_0)^2}}. & x\ge x_{cri}
 \end{array} \right. \eqno(12) $$
 Here $x_{cri}$, which satisfies the relation of ${(x_{cr}-x_0)^2\over
 r_p^2+(x_{cr}-x_0) ^2}={(R+x_{cri})^2 -(R+x_0)^2 \over (R+x_0)^2}$, is
 the critical height at which the horizon is just the area of the polar
 cap, and $R$ is the radius of the neutron star. Note that here we only
 consider
 the thermal emission from the ``hot spot'' at the polar caps, while 
 neglecting
 the contribution of the whole surface. This is because only very high
 temperatures (see Fig.2,3) can make the effect of two-photon annihilation
 become important. Such a high temperature is not likely for a cooling
 neutron star unless additional re-heating takes place in its polar caps.

 Using Eqs.[10,12], we can calculate the temperature-dependent two-photon
 annihilation mean free path. In the RS75's vacuum gap model, the gap height
 cannot exceed
 $$h_{max}={r_p \over \sqrt{2}}=1.0\times 10^4 P^{1/2} {\rm cm}. \eqno(13)$$
 Thus we regard
 $$l_{\gamma-\gamma,min}\le h_{max} \eqno(14)$$
 as the absolute criterion for occurrence of the two-photon annihilation
 process. Hence, we compare the temperature-dependent $l_{\gamma-\gamma,min}$
 with $h_{max}$ in Fig.2. A critical temperature $T_{cri}$ is thus defined,
 beyond which the two-photon process becomes important. From Fig.2 we find
 that, $l_{\gamma-\gamma,min}$ is very sensitive to the polar cap temperatures.
 When the temperature increases slowly near $T_{cri}$, the mean free 
 path may drop
 tremendously from infinity to less than $h_{max}$. For a pulsar period
 $P=0.1$s, $T_{6,cri}\sim 4$. This means that the temperature necessary to
 switch this $\gamma-\gamma$ process on is very high. In Fig.2, the three
 mean-free-path curves for different periods (curves 1,2,3) merge together
 eventually at some temperature, beyond which the mean free paths are less
 than the height of the ``hot corona'' $x_0$.

\begin{figure}
 %\picplace{6.0 cm}
 \centerline{\psfig{figure=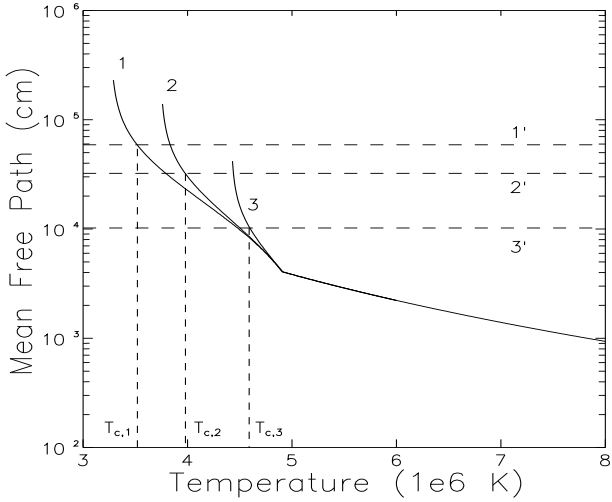,angle=0,height=7.0cm,width=9.0cm}}
 \caption[]{The temperature dependence of the shortest mean free paths of
 two-photon annihilation. Solid lines indicate the temperature-dependent
 mean free path of two-photon annihilation, while the dashed lines denote
 the corresponding maximum gap height. The numbers 1, 2, 3 indicate pulsar
 periods $P=0.03, 0.1, 1$s, respectively. We find that a critical temperature
 $T_c$
 clearly exists when the pulsar period is given.}

\end{figure} 

 The critical temperature ($T_{cri}$) only depends on the pulsar period $P$,
 which determines the size of the polar cap, and hence, determines the
 importance of this effect. However, the $T_{cri}$ variation is insensitive
 to $P$. Fig.3 plots the critical temperatures for two-photon annihilation
 for different pulsar periods. We find that $T_{6,cri}$ increases with 
 increasing period. The increasing rate is rapid for the fast-rotating pulsars,
 but becomes slower when the periods get longer. For most pulsars with period
 $P\sim 0.03-1.5s$, $T_{6,cri}$ ranges from 3.5-4.7.

\begin{figure}
 %\picplace{6.0 cm}
 \centerline{\psfig{figure=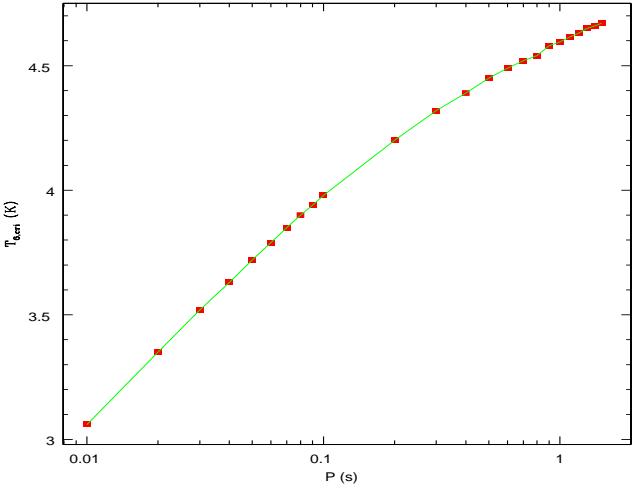,angle=0,height=7.0cm,width=9.0cm}}
 \caption[]{Period-dependence of the critical temperatures.}
\end{figure} 

 Now we see that the importance of the two-photon annihilation for forming 
 pair cascades in the pulsar polar cap depends completely on the temperatures
 of the polar caps. But can the polar caps be that hot?

 Theoretical analysis and recent X-ray observations do show that pulsars are
 likely to have a hot polar cap with temperatures of the order of $10^6K$.
 In the inner gap model, the maximum polar cap temperature can be achieved
 by assuming the rotation energy loss deposited on the conventional polar
 cap area
 $A=\pi r_p^2=6.6\times 10^8/P {\rm cm^2}$. This gives
 $$T_{max}=({\dot E_{rot}\over \sigma\pi r_p^2})^{1/4}\simeq 5.7\times
 10^6 \dot P_{-15}^{1/4} P^{-1/2} {\rm K}, \eqno(15)$$
 where $\sigma=5.67\times 10^{-5} {\rm erg\/ cm^{2}\/ K^{4}
 \/ s}$. Such high temperatures are indeed
 observed, e.g. $T_6\sim 1.5$ for PSR 0656+14 and $T_6\sim 3.7$ for
 1055-52 (Greiveldinger et al. 1996), $T_6\sim 5.14$ for PSR 1929+10 and
 $T_6\sim 5.70$ for PSR 0950+08 (Wang \& Halpern 1997). In view of this,
 we see that the two-photon process should play an non-negligible role in
 the pulsar polar cap physics.

\section{Comparing the $\gamma-\gamma$ and the $\gamma-B$ processes}

 The final pair production cascades are the consequence of competition
 among various processes. To judge the relative importance of these
 processes, we again use the ``shortest mean free path principle'' discussed
 in Sect.1. So it is necessary to compare the above-discussed $\gamma-\gamma$
 process with the conventional $\gamma-B$ process, which has been long
 regarded as the only important mechanism to form polar cap pair cascades.

 The basic difference between the two processes is that, the $\gamma-\gamma$
 mean free path strongly depends on polar cap temperatures, while the
 $\gamma-B$ mean free path is quite sensitive to the field line curvatures.
 Hence, a direct picture is: $\gamma-B$ cascades can develop easily near
 the rim of the polar cap, but not easily near the magnetic pole where the
 curvature is weak. The $\gamma-\gamma$ process, however, has the almost same
 probability to be developed all over the polar cap region. In the RS75 
 inner gap model, it is reasonable to assume that the gap potential only
 breaks down with a local spark where the pair-production avalanche takes
 place, but keeps unbroken in regions where pair cascade condition does
 not hold. This is because the pair flows cannot cross the field lines
 to fill the vacuum region other than the local spark, as the transverse
 energies of the secondary pairs are lost rapidly via synchrotron radiation.
 Thus a picture of gap sparking is: the gap grows at the speed of light, then
 sparks develop at the rim of the gap edge via the $\gamma-B$ process,
 but the gap height keeps growing near the pole until more
 $\gamma-B$ sparks take place.
 Finally, a $\gamma-\gamma$ induced central spark might be formed if the
 surface temperature is in excess of $T_{cri}$. While this picture is not
 presented quantitatively, we believe that it is at least
 qualitatively sound though the cascade near the rim might compensate for the
 lack of cascades in the central region. 

 Take a similar absolute criterion
 $$l_{\gamma-B,min}\le h_{max} \eqno(16)$$
 as Eq.[14], we examine how large is the area near the pole
 where the $\gamma-B$
 process fails. In the pure dipolar configuration, consider a field line whose
 ``root'' is at a colatitude of $\xi\theta_p$, where $\theta_p=r_p/R$ is the
 size of the polar cap and $0<\xi<1$ indicates the distance to the pole. Then
 the local field line curvature reads
 $$\rho={4\over 3}({\xi^{-2}\Omega R \over c})^{1/2}\simeq
 9.21\times 10^7\xi^{-1}P^{1/2}{\rm cm}. \eqno(17)$$
 The characteristic parameter
 $$\chi={E_\gamma \over 2m_0c^2}{B_\bot \over B_q}={E\over 2m_0c^2}{B \over
 B_q}{l_{\gamma-B}\over \rho} \eqno(18)$$
 to calculate the $\gamma-B$ mean free path (Erber 1966) is around 1/15 when
 $l_{\gamma-B}$ is of the order of $h_{max}$ (RS75). Thus the area where
 $\gamma-B$ fails is
 $$\xi\le 2.7\times 10^4({\chi\over 15^{-1}})P^{1/2}\dot P_{-15}^{-1/2}
 E_\gamma^{-1}{\rm (Mev)}. \eqno(19)$$
 Using Eqs.[2,3], and adopt
 $$\gamma=\gamma_{max}={e\over m_0c^2}{\Omega B\over c}h_{max}^2\simeq
 1.3\times 10^7 P^{-3/2}\dot P_{-15}^{1/2}, \eqno(20)$$
 we finally get
 $$\xi_{res}\le 0.09({\chi\over 15^{-1}})P^{3/2}\dot P_{-15}^{-3/2} \eqno(21)$$
 and
 $$\xi_{th}\le 6.6\times 10^{-7}({\chi\over 15^{-1}})P^{7/2}\dot P_{-15}^{-3/2}
 T_6^{-1}. \eqno(22)$$
 If the surface temperatures are higher than $T_{cri}$, Eqs.[21,22] just
 indicate the smallest area of the $\gamma-\gamma$ -induced central spark
 for the ``resonant ICS mode'' and the ``thermal ICS mode'', respectively
 (ZQLH97a, ZQH97b).

 A conventional polar gap spark usually occupy a region characterized by the
 gap height (RS75; Gil et al. 1997; Gil \& Cheng 1998, hereafter GC98). Using
 the ICS-induced gap parameters (see ZQH97b, their Eqs.[12,20]), we find a
 conventional $\gamma-B$ spark usually occupy a proportion
 $${h_{res}\over r_p}\simeq 0.075P^{1/3}\dot P_{-15}^{-1/2} \eqno(23)$$
 or
 $${h_{th}\over r_p}\simeq 1.9\times 10^{-2}P^{3/5}\dot P_{-15}^{-3/10}
 T_6^{-1/5} \eqno(24)$$
 of the polar cap. Comparing Eqs.[23,24] with Eqs.[21,22], we find that,
 if the $\gamma-B$ process is in the resonant mode, the central $\gamma-
 \gamma$-induced central spark is large enough to be compared with the
 conventional sparks, and hence, have a non-negligible effect in forming polar
 cap pair cascades. If the $\gamma-B$ process is in the thermal mode, however,
 the central $\gamma-\gamma$ spark is too small compared with the conventional
 ones, and hence, negligible.

\section{Possible connection with core emission}

 It is known that the radio emission beams of pulsars usually have two 
 kinds of components, namely the ``core'' and the ``conical'' ones (two cones, 
 Rankin 1983; or one cone, Lyne \& Manchester, 1988). The ``conical'' 
 components are a direct consequence of the hollow cone model, while the 
 ``core'' components, which can not interpreted directly by
 the conventional curvature radiation 
 model (e.g. Sturrock 1971; RS75), have stimulated many imagination
 (e.g. Beskin, Gurevich \& Istomin 1988; Wang, Wu \& Chen 1989; Weatherfall
 \& Eilek 1997; Qiao 1988, Qiao \& Lin 1998, hereafter QL98; Gil \& Snakowski
 1990). Among these models, the latter two require the inner gap sparking
 process, e.g. the pair production avalanches continuously developed in a
 vacuum gap in the polar cap region supposing that positive ions are bound
 in the neutron star surface (RS75). The inverse Compton scattering model
 of Qiao et al. (Qiao 1988; QL98; Xu \& Qiao 1998) discusses the pulsar radio
 observation properties (e.g. emission beams, polarization properties) as a
 consequence of the ICS of the secondary particles with the low-frequency wave
 produced by the inner gap sparking. This model can naturally results in the
 ``core'' and ``conical'' emission components of the pulsars, with their
 properties being interpreted successfully. The sparks of this model are
 supposed to be located at a rim in the polar cap near the last field lines,
 which is a natural consequence of the $\gamma-B$ pair production process in
 a standard dipolar configuration. In this model, the ``core'' emission does
 not require to a spark located in the central region of the polar cap (though
 it is better if there IS one), but is a consequence of the geometrical effect
 of the coherent ICS processes at lower altitudes. The model by Gil et al.,
 however, requires a strong central spark near the magnetic axis of the pulsar
 to account for the ``core'' emission, with some other weak ones drifting
 around the central one to account for the ``conical'' emissions (e.g. Gil \&
 Kijak 1992). Recently, they (Gil et al. 1997; GC98) found that it is
 reasonable to assume that the sparks are located all over the polar cap
 region, with characteristic dimension as well as the typical
 distance between neighbours approximately equal to the height
 of the gap. This picture is supported by observations, since those
 pulsars showing complex profiles (e.g. cT, Q and M profiles, Rankin 1983)
 all lie in the region in the $P-\dot P$ diagram where the parameter
 $a=r_p/h\simeq 4$.

 However, in the conventional gap sparking model with standard dipolar
 magnetic field configuration, it is quite difficult to get a central spark
 near the pole due to the ``weak curvature problem''. Although this problem
 might be solved by invoking multipolar magnetic field components (see, e.g.
 GC98), it is still uncertain since we don't know the real field
 configurations in the neutron star vicinity. A possible way out is to
 invoke the induced electric field as the dominant mechanism for the
 $\gamma$-photon absorption as suggested by Daugherty \& Lerche (1975),
 but Zheng, Zhang \& Qiao (1998) recently argued that this electric field
 absorption picture is unfortunately wrong and should be abandoned since the
 incident angles of the $\gamma$-photons are misused.

 The $\gamma-\gamma$ process discussed in this paper seems to present another
 possible mechanism to account for the central sparks in the pulsar polar
 caps. 
 Furthermore, it is interesting that the $\gamma-\gamma$-induced central
 sparks seem to have some properties relevant to the pulsar ``core'' emission.
 These evidence include:

 1. ``Core'' emission components are usually not observed in the pulsars near
 the death line (Rankin 1990). This is understandable in our picture, since the
 region near the death line is just where the thermal ICS mode dominates the
 resonant ICS mode (ZQLH97a). The two hot-polar-cap old pulsars PSR 1929+10
 and PSR 0950+08 (Wang \& Halpern) should be in the thermal ICS mode 
 according to our analysis. Actually, they do not show ``core'' components
 in their pulse profiles (Rankin 1990, 1993; GC98). 

 2. Pulsars with ``core'' emission components have relative shorter periods
 on average (Rankin 1990). It is natural in our picture, since shorter periods
 result in a lower threshold for the two-photon process (see Fig.3).
 In this picture, millisecond pulsars should have the tendency to show the
 ``core'' feature. Though polarization data of millisecond pulsars are not
 sufficient to draw this conclusion, some millisecond pulsars do show central
 components in their pulse profiles (Han 1998, private communication). We
 hope that this prediction can test as the polarization data accumulate.

 3. For the $\gamma-\gamma$ process, the secondary pairs are more energetic
 than those from the $\gamma-B$ one. This is because the $\gamma-\gamma$
 secondaries inherit the energy of the high energy photons 
 ($\gamma_{\pm}\sim E/(2m_0c^2$), while the $\gamma-B$ secondaries 
 rapidly lose their perpendicular energies via synchrotron radiation, so
 that only the parallel energy components remains (see ZQH97b). This is just
 the requirement of the ``core emission'' models, since higher energies
 are required for the particles along the central nearly-straight field
 lines both in the ICS (QL98) or CR (e.g. RS75) models in order to account
 for the emission with same frequencies.

 4. The two photon annihilation cross section is greatly reduced with respect
 to $\epsilon E (1-\mu_c)= 2(m_0c^2)^2$. This will results in steep spectra of
 the secondary particles, and hence, the steeper spectra observed in pulsar
 ``core'' emissions (Rankin 1983, 1990).

 However, although we have mentioned evidences for the possible
 connection of the $\gamma-\gamma$-induced central spark with pulsar core
 emission, it is difficult to realize a large central spark in practice.
 This is due to the high critical temperature ($T_{6,cri}\sim 4$) for the
 two-photon annihilation. As discussed in Sect.2, it is not difficult to
 realize a hot polar cap with $T_6\sim 5$. The difficulty is that it is hard
 to find a pulsar with polar cap temperature $T_6\sim 4$ but is still in
 the ``resonant ICS mode''. As discussed in ZQLH97a, there is still another
 critical temperature $T^{\prime}_{cri}$ to distinguish the two ICS
 $\gamma-B$-induced gap modes, which is also around $(3-4)\times 10^6K$.
 So it is quite likely that, when the polar cap is hot enough to switch
 $\gamma-\gamma$ sparks, the $\gamma-B$ sparks are dominated by the
 thermal ICS mode, which produces much more energetic $\gamma$-rays. In
 this case, the $\gamma-\gamma$-induced central spark will be too small
 to play an important role (see Eqs.[22]). But in contrast with $T_{cri}$,
 which is a strong constraint since $\gamma-\gamma$ mean free path is quite
 sensitive around it, $T^{\prime}_{cri}$ is rather uncertain since we do
 not know how violent the pair avalanche is. Different assumptions can change
 $T^{\prime}_{cri}$ by a factor of 2-3 (see detailed discussion in ZQLH97a).
 So there is still some room for the existence of a large-enough $\gamma-
 \gamma$-induced central spark when the $\gamma-B$ process is dominated by
 the resonant ICS mode. Furthermore, Eqs.[21,22] just show the smallest area
 of the central sparks. The central sparks can be even larger for a higher
 temperature since the $\gamma-\gamma$ mean free path will be much smaller.
 With all these in mind, we conclude that the question whether the
 $\gamma-\gamma$ process can form an important core-emission-connected
 central spark is still open.

\section{Summary and discussions}

 We have discussed the possible importance of the two-photon annihilation in
 the pulsar polar gap sparking process. We found that this process may not be
 negligible when the polar cap temperature $T\ge T_{cri}\sim (3.5-4.7)\times
 10^6$K. Compared with the $\gamma-B$ process, although this two-photon
 process is not important when the $\gamma-B$ process is dominated by the
 thermal ICS mode, it can be an important mechanism to form a central spark
 near the pole if the resonant ICS mode dominates. The uncertainty to
 distinguish the two ICS modes does not allow or rule out this possibility.
 If so, then this mechanism could be a possible interpretation
 to GC98's spark model.

 There are some factors which might more or less influence the importance of
 the two-photon effect. First, we have used Eq.[10] to calculate $l_{\gamma-
 \gamma,min}$, in which Max$f[\nu]$ is used generally. This might have
 enhanced the importance of this effect a bit since the value $\nu$ at
 which Max$f[\nu]$ is achieved is not exactly aligned for different $\mu_c$.
 But this variation is not large, so it does not severely influence the value
 of $T_{cri}$. Second, within the neutron star magnetosphere, the temperature
 is higher near the surface (Pavlov et al. 1995). This effect will enhance the
 importance of the two-photon process, since we only observe a ``cooler''
 emission from infinity. Third, we have adopted a blackbody-like thermal
 photon field to do the calculation. This should be modified since neutron
 stars are not perfect blackbodies due to strong magnetic fields in the
 star vicinity (Pavlov et al. 1995). But the modification of the spectral shape
 will not change much the characteristic energy and the number
 density of the soft
 photon fields where they are the key parameters to determine the two-photon
 annihilation mean free path (see Eq.[5]). So the results for the
 mean free paths and the critical temperatures will not change much.
 Fourth, if strong multipole magnetic components with stronger curvatures
 do exist at the polar caps, the $\gamma-\gamma$ process will be further
 suppressed, so that the $\gamma-\gamma$-induced central spark would be 
 smaller. Finally, we did not consider the influence of the strong magnetic
 fields on the two-photon annihilation cross section, which might yield some 
 non-negligible modifications (see e.g. Burns \& Harding 1984; 
 Harding 1998, private communication). Both the cross sections of the two
 photon annihilation and its inverse process pair annihilation
 ($e^{+}+e^{-}\rightarrow \gamma+\gamma^{\prime}$) are reduced with the
 presence of strong magnetic fields (Daugherty \& Bussard 1980;
 Baring \& Harding 1992; Kozlenkov \& Mitrofanov 1987), so that the critical
 temperature $T_{cri}$ may be higher. Further analysis is required to
 present a quantitative analysis.

 Our discussion throughout this paper is within the framework of the
 RS75 vacuum gap model. There are some observational evidences for
 unsteady flows of particles from the pulsar inner magnetospheres, such as
 microstructures, drifting sub-pulses, etc. Though a binding energy problem
 has been long recognized, there are still some theoretical approaches which
 support the RS-type gap (see a detailed discussion in GC98, Xu \& Qiao
 1998b). Furthermore, the vacuum gap model (RS75, ZQH97b) gives at least
 order-of-magnitude parameters for the inner accelerators, such as the
 heights, parallel electric fields, the potential drops, etc. It is natural
 to notice that the conclusion of this paper can also be extended to the
 steady space-charge-limited-flow model (e.g. Arons \& Scharlemann 1979;
 Arons 1983; Muslimov \& Tsygan 1992; Muslimov \& Harding 1997; HM98). By
 taking the two-photon process into account, the ``slot gap'' (Arons 1983)
 will not be that ``slot'' near the pole if the polar cap temperature is
 hot enough ($T>T_{cri}$), since a pair formation front (PFF) will be formed
 there by the $\gamma-\gamma$ process.

 We did not consider the anisotropy between the outgoing and back-flow
 particles. A qualitative discussion can be performed. As pointed out
 by HM98, the backflow ICS will produce more energetic photons, so that
 the $\gamma-B$ process is enhanced. The $\gamma-\gamma$ process, however,
 will be also enhanced since head-on collisions are dominant.
 Thus, the relative importance of the two processes might not be much changed.

\acknowledgements{The authors thank the referee C. D. Dermer for his important
 comments and suggestions and careful correction of the language problems.
 We are also grateful to Alice Harding for her careful reading the manuscript
 and insightful comments and suggestions, to Janusz Gil and J.L. Han for their
 useful comments, to R.X. Xu, J.F. Liu, B.H. Hong, Z. Zheng for helpful
 discussions, and to Z. Zheng and B.H. Hong for technique assistence.
 This work is supported by the NNSF of
 China, the Climbing Project of China, and the Project Supported by Doctoral
 Program Foundation of Institution of Higher Education in China. BZ
 acknowledges supports from China Postdoctoral Science Foundation.}


\begin{thebibliography}{}

  \bibitem{} Arons, J. 1983, ApJ, 266, 215
  \bibitem{} Arons, J., \& Scharlemann, E.T. 1979, ApJ, 231, 854
  \bibitem{} Baring, M.G., \& Harding, A.K. 1992, Proc. of the 2nd GRO
	     Science Workshop, ed. C.R. Schrader, N. Gehrels \& B. Dennis
             (NASA CP-3137), 245
  \bibitem{} Becker, W., \& Tr\"umper, J. 1997, A\&A, 326, 682
  \bibitem{} Beskin, V.S., Gurevich, A.V., \& Istomin, Y.N. 1988, ApSS, 
	     146, 205
  \bibitem{} Burns, M.L., \& Harding, A.K. 1984, ApJ, 285, 747
  \bibitem{} Chang, H.K. 1995, A\&A, 301, 456
  \bibitem{} Daugherty, J.K., \& Bussard, R.W. 1980, ApJ, 238, 296
  \bibitem{} Daugherty, J.K., \& Harding, A.K. 1989, ApJ, 336, 861
  \bibitem{} Daugherty, J.K., \& Lerche, I. 1975, ApSS, 38, 437
  \bibitem{} Dermer, C.D. 1990, ApJ, 360, 214
  \bibitem{} Erber, T. 1966, Rev. Mod. Phys. 38, 626
  \bibitem{} Gil, J., \& Cheng, K.S. 1998, MNRAS, submitted: GC98
  \bibitem{} Gil, J., \& Kijak, J. 1992, A\&A, 256, 477
  \bibitem{} Gil, J., Krawczyk, A., \& Melikidze, G. 1997, In: Mathematics
	     of Gravitation, Banach Center Publications, Vol. 41, 239
  \bibitem{} Gil, J., Snakowski, J.K. 1990, A\&A, 234, 237
  \bibitem{} Greiveldinger, C., et al. 1996, ApJ, 465, L35
  \bibitem{} Gould, R.J., Schr\'{e}der, G.P. 1967, Phys. Rev. 155, 1404: 
	     GS67
  \bibitem{} Harding, A.K., \& Muslimov, A.G. 1998, ApJ, in press: HM98
  \bibitem{} Kozlenkov, A.A. \& Mitrofanov, I.G. 1987, Sov. Phys-JETP,
             64, 1173
  \bibitem{} Luo, Q. 1996, ApJ, 468, 338
  \bibitem{} Lyne, A.G., \& Manchester, R. N. 1988, MNRAS, 234, 477
  \bibitem{} Muslimov, A.G., \& Harding, A.K. 1997, ApJ, 485, 735
  \bibitem{} Muslimov, A.G., \& Tsygan, A.I. 1992, MNRAS, 255, 61
  \bibitem{} \"Ogelman, H. 1995, in: The Lives of the Neutron Star, ed. 
	     M.A. Alpar, \"U. Kiziloglu, \& J. van Paradijs (NATO ASI Ser. 
	     C, 450) (Dordreicht: Kluwer), 101
  \bibitem{} Pavlov, G.G., Shibanov, Yu.A., Zalvin, V.E., \& Meyer, R.D. 
	     1995, in: The Lives of the Neutron Star, ed. 
	     M.A. Alpar, \"U. Kiziloglu, \& J. van Paradijs (NATO ASI Ser. 
	     C, 450) (Dordreicht: Kluwer), 71
  \bibitem{} Qiao, G.J. 1988, Vistas in Astronomy, 31, 393
  \bibitem{} Qiao, G.J., \& Lin, W.P. 1998, A\&A, 333, 172: QL98
  \bibitem{} Qiao, G.J., \& Zhang, B. 1996, A\&A, 306, L5
  \bibitem{} Rankin, J.M. 1983, ApJ, 274, 333
  \bibitem{} Rankin, J.M. 1990, ApJ, 352, 247
  \bibitem{} Rankin, J.M. 1993, ApJ, 405, 285
  \bibitem{} Ruderman, M.A., Sutherland, P.G. 1975, ApJ, 196, 51: RS75
  \bibitem{} Sturner, S.J. 1995, ApJ, 446, 292
  \bibitem{} Sturner, S.J., Dermer, C.D. \& Michel, F.C. 1995, ApJ, 445, 736
  \bibitem{} Sturrock, P.A. 1971, ApJ, 164, 529
  \bibitem{} Wang, D.Y., Wu, X.J., \& Chen, H. 1989, ApSS, 116, 217
  \bibitem{} Wang, F.Y.-H., \& Halpern, J.P. 1997, ApJ, 482, L159
  \bibitem{} Weatherfall, J.M., Eilek, J.A. 1997, ApJ, 474, 407
  \bibitem{} Xia, X.Y., Qiao, G.J., Wu, X.J., \& Hou, Y.Q. 1985, A\&A, 
	     152, 93
  \bibitem{} Xu, R.X., \& Qiao, G.J. 1998a, ApJ, submitted
  \bibitem{} Xu, R.X., \& Qiao, G.J. 1998b, ApJ, submitted (astro-ph/9804278)
  \bibitem{} Zhang, B., \& Qiao, G.J. 1996, A\&A, 310, 135: ZQ96
  \bibitem{} Zhang, B., Qiao, G.J., Lin, W.P., \& Han, J.L. 1997a, ApJ, 
	     478, 313: ZQLH97a
  \bibitem{} Zhang, B., Qiao, G.J., \& Han, J.L. 1997b, ApJ, 491, 891: 
	     ZQH97b
  \bibitem{} Zheng, Z., Zhang, B., \& Qiao, G.J. 1997, A\&A, 334, L49 





\end{thebibliography}
\end{document}